\documentclass[prd,nofootinbib,twocolumn,superscriptaddress,preprintnumbers,balancelastpage]{revtex4-1}

\usepackage{graphicx}
\graphicspath{{images/}}
\usepackage{color}
\usepackage{hyperref}
\hypersetup{
    colorlinks = true,
    citecolor  = red,
	linkcolor  = blue
}
\usepackage{verbatim}
\usepackage{fontawesome} 
\usepackage{xspace} 
\usepackage[normalem]{ulem} 

\usepackage{amsmath, amssymb}
\usepackage{halloweenmath}

\newcommand{\dd}{\text{d}}
\newcommand\order[1]{\mathcal O(#1)}

\definecolor{yscolor}{rgb}{0.1, 0.6, 0.1}
\definecolor{yscolorp}{rgb}{0.0, 0.6, 0.4}

\newcommand{\dhis}{\texttt{DarkHistory}\xspace}
\newcommand{\Npropg}{{\mathbf{N}_\text{prop}^\gamma}}
\newcommand{\Nlowg}{{\mathbf{N}_\text{low}^\gamma}}
\newcommand{\Nlowe}{{\mathbf{N}_\text{low}^e}}
\newcommand{\Ec}{{E^\text{high}_c}}
\newcommand{\Pbg}{{\mathbf{\overline{P}}_\gamma}}
\newcommand{\Dbg}{{\mathbf{\overline{D}}_\gamma}}
\newcommand{\Dbe}{{\mathbf{\overline{D}}_e}}
\newcommand{\Dbc}{{\mathbf{\overline{D}}^\text{high}_c}}
\newcommand{\Eg}{{\mathcal{E}^\gamma}}
\newcommand{\Ee}{{\mathcal{E}^e}}
\newcommand{\Ei}{{\mathcal{E}_i}}
\newcommand{\Sbg}{{\mathbf{\overline{S}}_{\gamma,c}}}

\newcommand{\ourtitle}{
    Modeling early-universe energy injection with Dense Neural Networks
}

\begin{document}

\preprint{MIT-CTP/5449}
\title{\ourtitle \vspace*{-0.3cm}}

\author{Yitian Sun}
\email{yitians@mit.edu}
\affiliation{Center for Theoretical Physics, Massachusetts Institute of Technology, Cambridge, MA 02139, USA}
\affiliation{The NSF AI Institute for Artificial Intelligence and Fundamental Interactions}

\author{Tracy R. Slatyer}
\email{tslatyer@mit.edu}
\affiliation{Center for Theoretical Physics, Massachusetts Institute of Technology, Cambridge, MA 02139, USA}
\affiliation{The NSF AI Institute for Artificial Intelligence and Fundamental Interactions} 

\begin{abstract} \noindent
We show that Dense Neural Networks can be used to accurately model the cooling of high-energy particles in the early universe, in the context of the public code package \texttt{DarkHistory}. \texttt{DarkHistory} self-consistently computes the temperature and ionization history of the early universe in the presence of exotic energy injections, such as might arise from the annihilation or decay of dark matter. The original version of \texttt{DarkHistory} uses large pre-computed transfer function tables to evolve photon and electron spectra in redshift steps, which require a significant amount of memory and storage space. We present a light version of \texttt{DarkHistory} that makes use of simple Dense Neural Networks to store and interpolate the transfer functions, which performs well on small computers without heavy memory or storage usage. This method anticipates future expansion with additional parametric dependence in the transfer functions without requiring exponentially larger data tables. \href{https://github.com/hongwanliu/DarkHistory}{\faGithub}
\end{abstract}

\maketitle

\section{Introduction}

\noindent
Dark matter (DM) constitutes 84\% of the matter content in the universe \cite{aghanim2020planck} and plays an important role in the evolution of the early universe. It has so far eluded detection in all channels other than gravitational interactions. DM annihilation or decay could inject energy in the form of Standard Model particles, modifying the temperature and ionization of the intergalactic medium (IGM) and the anisotropies of the cosmic microwave background (CMB); studies of these observables have placed strong constraints on such energy injections (e.g. \cite{adams1998cosmic, chen2004particle, padmanabhan2005detecting, zhang2006impacts, slatyer2009cmb, kanzaki2010effects, slatyer2013energy, galli2013systematic,Slatyer:2015jla,Slatyer:2015kla,liu2016contributions, slatyer2017general, poulin2017cosmological, liu2021lyman}).

\dhis \cite{liu2020darkhistory} is a \texttt{Python} package developed to calculate the evolution of the IGM temperature and ionization in the early universe in the presence of such exotic energy injections. For an injected spectrum of Standard Model (SM) particles, it calculates the particle cascade by computing
(1) the production of photons and electrons/positrons by the decay of the originally injected SM particles;
(2) the subsequent secondary particle cascade and energy deposition arising from this exotic injection of photons/electrons/positrons, due to interaction with the IGM and the photon bath;
(3) modifications to the IGM's temperature and ionization from the secondary particles and their energy deposition, using a simple Three-Level Atom (TLA) model.

These calculations are carried out in redshift steps starting prior to recombination (at redshift $1+z=3000$ by default) and ending well after reionization near the present day (redshift $1+z=4$). In particular, the particle cascade in step (2) is evaluated using precomputed transfer functions, which are matrices that take an input spectrum and output the spectrum of secondary particles (for a given redshift step). \dhis includes the backreaction effects of changes to the ionization level of matter, which means the transfer functions themselves are functions of the gas ionization levels, as well as redshift. In the previous version of \dhis, this dependence is realized by interpolating tables of transfer function matrices on a grid of values for the hydrogen and helium ionization fractions, as well as a grid of redshift values.

\begin{figure}[ht]
\includegraphics[clip, width=0.4\textwidth]{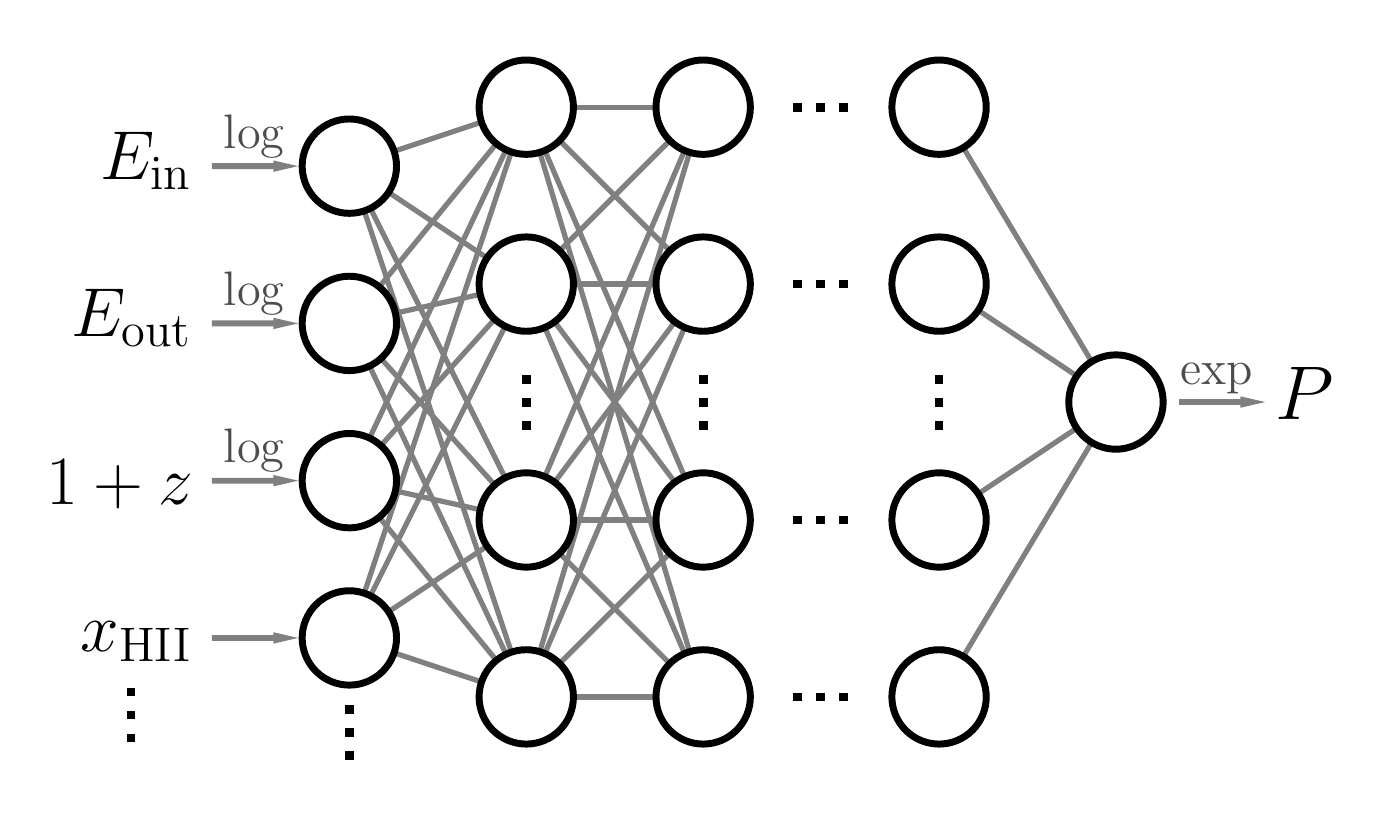}
\caption{\textbf{Model schematics of a Dense Neural Network transfer function.} A DNN takes in (the logarithm of) the input and output photon/electron energy and necessary physical parameters including the redshift $z$, ionized hydrogen fraction $x_\text{HII}$, and singly ionized helium fraction $x_\text{HeII}$, and outputs (the logarithm of) the transfer function value $P$. The transfer function matrix acting on a given discretized spectrum is then obtained by evaluating the DNN on the given energy abscissa.}
\label{fig:model}
\end{figure}

At around $1.5$~Gb per table with 12 tables around this size, the transfer functions take up significant storage space as well as memory during runtime, since they are all loaded in a standard run. They will also be difficult to scale up to include additional parametric dependence, as the expected size scales exponentially with the number of added parameters. Intuitively, storing the transfer functions as tables is an over-representation of the information content within, since the transfer functions, despite not being smooth globally, can be divided into multiple regions that are relatively smooth, each of which could plausibly be fitted with analytical functions. However, in practice, finding such a solution is quite non-trivial, and even if a solution was found by \emph{ad hoc} methods, it would be quite specific to individual transfer functions and potentially difficult to maintain under changes to the code.
In this work, we present a general solution to these issues by replacing the transfer function tables with trained Dense Neural Networks.

Recently, machine learning and especially Neural Networks (NNs) has seen many applications in high energy physics and astrophysics \cite{lecun2015deep, bourilkov2019machine, radovic2018machine, albertsson2018machine}. NNs are often used to capture highly nonlinear and complex relations between inputs and outputs, and in particular, Dense Neural Networks (DNNs), also known as fully connected Neural Networks, are a class of general function approximators given sufficient neuron numbers \cite{Goodfellow-et-al-2016}.

\begin{figure*}[ht]
\includegraphics[clip, width=0.7\textwidth]{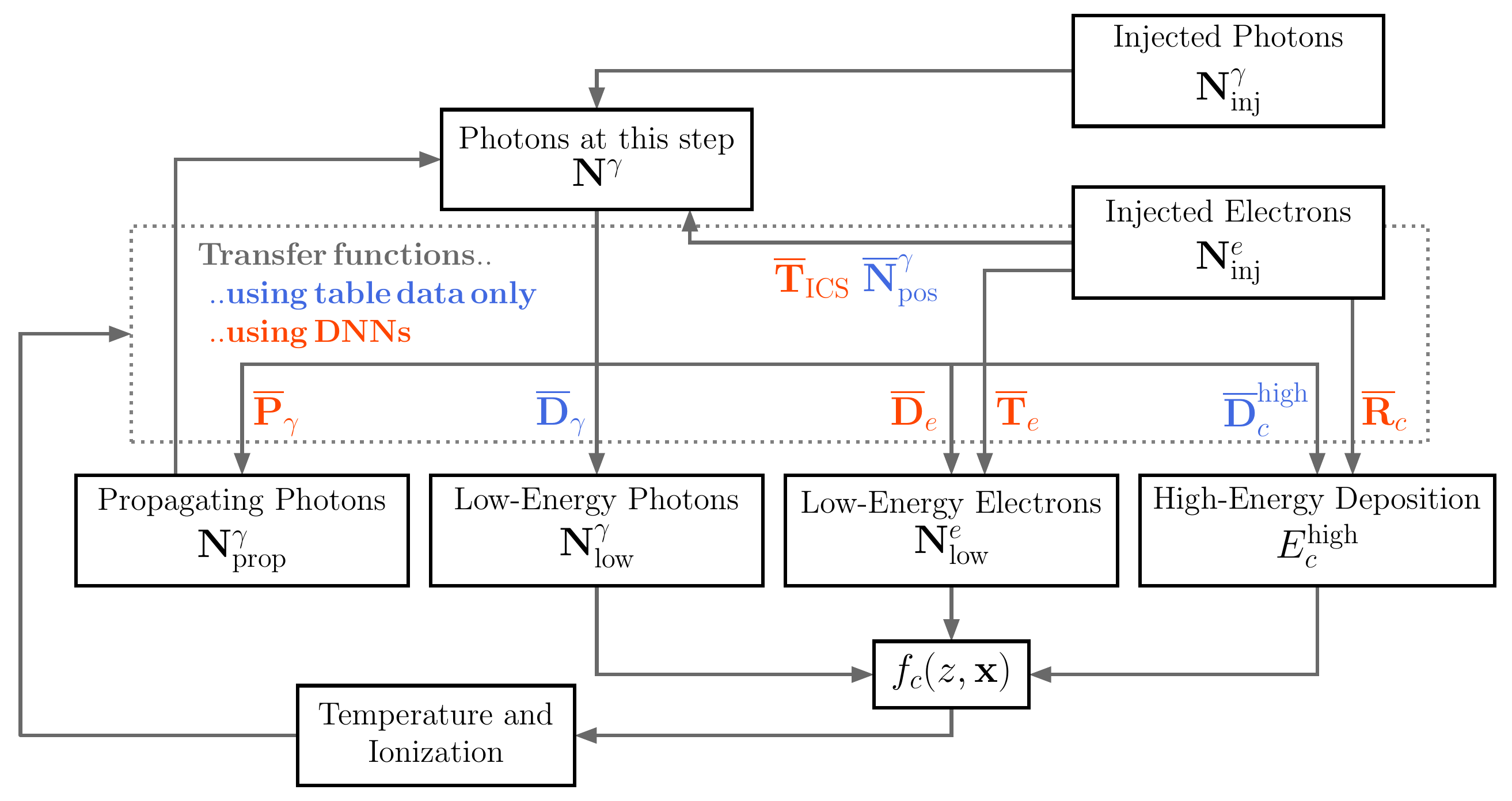}
\caption{\textbf{\dhis flow chart and transfer functions.} Starting with injected photons and electrons, this flow chart illustrates the work flow of \dhis in the evolution of each redshift step. The boxed quantities are particle spectra and the arrows indicate transfer functions, which act on the particle spectra. The orange transfer functions are generated or reconstructed using DNNs, while the blue ones only use tabulated data. (Modified from Fig.~1 in Ref.~\cite{liu2020darkhistory}.) }
\label{fig:dhtfs}
\end{figure*}

In the updated version of \dhis we present in this work, we use lightweight DNNs to store and automatically interpolate \dhis's transfer functions. A transfer function is essentially a multi-dimensional table, with 2 dimensions corresponding to the input and output particle energy, and the rest corresponding to physical parameters which in this case will be a subset of redshift $z$, ionized hydrogen fraction $x_\text{HII}\equiv n_\text{HII}/n_\text{H}$, and singly ionized helium fraction $x_\text{HeII}\equiv n_\text{HeII}/n_\text{H}$. (As a matter of convenience, we define the singly ionized helium fraction with the \emph{hydrogen} number density as the denominator, so that $x_\text{HeII}$ and $x_\text{HII}$ can be easily summed.) The DNN-based transfer functions are shown in schematic form in Fig.~\ref{fig:model}. We let a DNN take in all of the relevant parameters on equal footing and predict (the natural logarithm of) the transfer function value $P$.

Using DNNs as transfer functions has several benefits:

\begin{itemize}
    \item The DNNs we use are smaller in size compared to the stored transfer function tables, by a factor of $\sim$400.
    \item The computed matter temperature history and ionization history
    match those calculated using the transfer function tables to within a few percent relative difference (with sub-percent relative differences in regions when the species in question are more than 10\% ionized), while the spectral distortion due to upscattered CMB photons (see Sec.~\ref{sec:tf} for precise definition) matches to below the 10 percent level. These errors are small compared to current experimental uncertainties.
    \item We expect the DNNs to have improved scaling in size when additional parameters are added compared to the original tables. (Including a smooth dependence on an additional parameter might result in a $\order1$ increase in the DNN neuron number to reach similar accuracy due to increased information content. On the other hand, adding an additional dimension to a data table would increase its size by a multiplicative factor of the number of bins in the new parameter.)
    \item The DNNs automatically interpolate to any the physical parameter values and input/output particle energies within the trained range. This allows the use of flexible binning in \dhis, and will also allow us to perform interpolation on sparse training data \cite{duchi2011adaptive, duchi2013estimation}. The latter may become necessary in future extensions of \dhis, when probing dependence on an increasing number of physical parameters and generating dense grids of training data becomes computationally expensive.
    \item The DNNs predict transfer functions quickly, taking up a similar amount of time to the rest of the evolution routine for injected particles. Thus compared to retrieving tabular data from memory on a personal computer, the use of DNNs results in only a $\order1$ increase in total runtime.
    \item Open source building and training tools for NNs and especially simple architectures like DNNs are readily available. (In this work we use Tensorflow 2.0 \cite{tensorflow2015-whitepaper} with Keras \cite{chollet2015keras}.)
    
\end{itemize}

In Sec.~\ref{sec:tf}, we introduce \dhis and the roles of transfer functions. In Sec.~\ref{sec:nntf}, we detail the training and implementation of the DNN transfer functions. In Sec.~\ref{sec:performance} we present test runs and discuss the performance of DNN transfer functions compared to baseline \dhis. Finally, in Sec.~\ref{sec:conclusion} we summarize our results, briefly discuss other possible approaches, and outline some ideas for future expansion with DNN transfer functions in \dhis.

\section{Transfer functions in DarkHistory}
\label{sec:tf}

To better illustrate the role of transfer functions, we briefly introduce the procedure followed in \dhis and sketched in Fig.~\ref{fig:dhtfs}, which is modified from a flow chart in Ref.~\cite{liu2020darkhistory}. In Fig.~\ref{fig:dhtfs}, boxed quantities represent particle spectra, and arrows represent transfer functions, which are functions acting on spectra. \dhis stores the free streaming photon spectrum, the IGM temperature, and the IGM's ionized hydrogen fraction and singly ionized helium fraction at each redshift step (these quantities are assumed to be homogeneous). For each redshift step, \dhis:
\begin{enumerate}
    \item Converts injected SM particles at that redshift to injected photons $\mathbf{N}^\gamma_\text{inj}$ and electron/positrons $\mathbf{N}^e_\text{inj}$ (hereafter referred collectively as electrons). In \dhis, high energy ($>3$~keV) positrons are treated as electrons since their dominant energy-loss process, Inverse Compton Scattering (ICS) on the CMB, does not depend on the particle charge.  Lower-energy positrons are assumed to annihilate and the resulting photon spectrum is tracked; their kinetic energy is approximated as following the same pattern of energy deposition as that of the electrons (see Ref.~\cite{liu2020darkhistory} for a more in-depth discussion).
    \item Computes any injected electron spectrum's energy deposition into ionization, excitation, or heating by applying the transfer function $\overline{\mathbf{R}}_c$. Computes secondary photon and electron spectra produced from injected electrons due to ICS, positronium formation and decay, and atomic processes by applying the ICS transfer function $\overline{\mathbf{T}}_\text{ICS}$ and secondary electron transfer function $\overline{\mathbf{T}}_e$, and evaluating the spectrum of gamma rays produced from positron annihilation  $\overline{\mathbf{N}}^\gamma_\text{pos}$. These secondary photon spectra are added to the spectrum of photons $\mathbf{N}^\gamma$ propagated from the previous step, plus any injected photon spectrum.
    \item Computes the secondary particles and energy deposition produced by a propagating photon spectrum $\mathbf{N}^\gamma$, due to a variety of processes, including photon-photon scattering, Compton scattering, pair production, photoionization, and redshifting. The production of secondary electrons/positrons in the same redshift step, and their subsequent production of photons via ICS on the CMB or positron annihilation, are also included. The propagating photon spectrum for the next redshift step is obtained by applying the high energy photon transfer function $\overline{\mathbf{P}}_\gamma$ to $\mathbf{N}^\gamma$. The low energy photon spectrum, which stores photons below 3 keV that either photoionize within the redshift step or lie below 13.6 eV, is obtained by applying the low energy photon transfer function $\overline{\mathbf{D}}_\gamma$.
    The low energy electron spectrum, which stores electrons with kinetic energy below 3 keV where atomic cooling dominates ICS and is treated separately in the electron cooling module, is obtained by applying the low energy electron transfer function $\overline{\mathbf{D}}_e$. Finally, the photon's energy deposition into ionization, excitation, and heating is obtained by applying the high energy deposition transfer function $\overline{\mathbf{D}}_c^\text{high}$.
    \item Computes the change to IGM temperature and ionization by first calculating the energy deposition fraction $f_c$'s, from the low-energy electron/photon spectra and direct energy deposition by higher-energy particles, and then performing the TLA integration (see Ref.~\cite{liu2020darkhistory} for details).
\end{enumerate}

In general, transfer functions from an input spectrum to another spectrum (such as $\overline{\mathbf{T}}_\text{ICS}$, $\overline{\mathbf{P}}_\gamma$) take up much more space than those outputting an energy value (such as $\overline{\mathbf{D}}_c^\text{high}$) or those that are diagonal (such as $\overline{\mathbf{D}}_\gamma$). For this version of \dhis, we focus on replacing the largest spectral transfer functions with DNNs, but similar procedures can be applied to the lower-dimension transfer functions in the future. In the following we introduce the two major types of transfer functions we will replace in more detail.

\subsection{ICS transfer functions}

As discussed above, the ICS transfer functions describe the spectrum of scattered photons from the complete cooling of injected electrons. Unlike the transfer functions applied to the photon spectrum, the ICS transfer functions $\overline{\mathbf{T}}_\text{ICS}$, $\overline{\mathbf{T}}_e$ and $\overline{\mathbf{R}}_c$ are not directly interpolated from tables, but reconstructed from reference ICS transfer function tables. Sec.~III.D and Appendix A of Ref.~\cite{liu2020darkhistory} describe in detail how this is achieved, and we only provide a simple summary here: since an electron quickly deposits all of its energy in one of our redshift steps, in order to obtain the total secondary spectrum or energy deposition by an electron, we need to consider multiple interaction events (via ICS and atomic processes). This can be done recursively: one can reconstruct the full ICS secondary spectrum and energy deposition for an electron of energy $E$ knowing the same information for all electrons with $E'<E$. With discretized energy abscissa, the full ICS-induced photon spectrum and energy output can be solved recursively from the lowest energy bin. As a result, at each redshift, one can solve for the full electron ICS transfer functions using transfer functions describing a single ICS scattering event (as well as functions describing the interaction rates due to atomic processes).

The transfer functions for a single ICS scattering event on the CMB have simple (approximate) scaling relations with respect to the temperature of the CMB $T$, as described in detail in Appendix A of Ref.~\cite{liu2020darkhistory}. As such one can derive ICS transfer functions at different redshifts from a single transfer function at a fixed redshift, assuming the CMB is the dominant radiation background (in \dhis, $1+z=400$ is used). These reference transfer functions are interpolated from tables \texttt{ics\_thomson}, \texttt{ics\_rel}, and \texttt{ics\_engloss}, corresponding respectively to transfer functions for the secondary photon spectra of nonrelativistic electrons and relativistic electrons, and the relativistic electron energy loss in a single ICS scattering event on the CMB. It is these tables that we will fit with DNNs.

\subsection{Photon transfer function}
\label{subsec:photontf}

The high energy photon transfer function $\overline{\mathbf{P}}_\gamma$, low energy photon transfer function $\overline{\mathbf{D}}_\gamma$, and low energy electron transfer function $\overline{\mathbf{D}}_e$ are interpolated from corresponding tables \texttt{highengphot}, \texttt{lowengphot}, and \texttt{lowengelec}. They in general depend on the CMB temperature through redshift and matter ionization levels (ionized hydrogen fraction and singly ionized helium fraction). For each combination of these physical parameters, \texttt{lowengphot} can be represented as 1-D arrays (with the one dimension being input/output energy) and is a factor of 500 smaller than the other transfer functions, so it is at present not replaced with a DNN. We also found that the numerical calculation of Compton scattering used in the previous version of \dhis was inaccurate in some parts of parameter space (in particular populating kinematically forbidden regions), and so we have updated the relevant tabular transfer functions to ensure sufficient accuracy.\footnote{This issue primarily affected secondary electrons in the 10 eV to 3 keV range produced by Compton scattering of photons with energies between 100 eV and 10s of keV; the effects on observable quantities were very small for all cases we checked. The updated tables on \href{https://doi.org/10.5281/zenodo.6819281}{Zenodo} incorporate this correction.} We now describe some special features of the photon transfer functions:

\begin{figure*}[ht]
\includegraphics[clip, width=\textwidth]{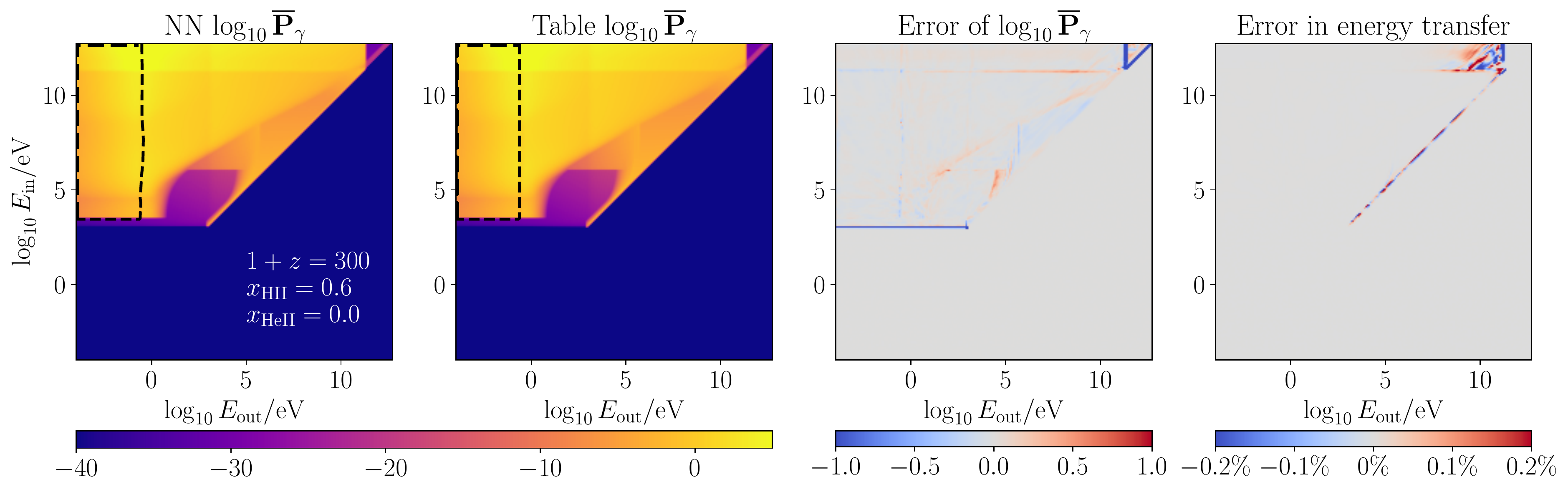}
\caption{\textbf{Comparison of DNN and tabulated transfer functions.} The first two panels show the high energy photon transfer functions generated at redshift $1+z=300$, ionized hydrogen fraction $x_\text{HII}=0.6$ (and singly ionized helium fraction $x_\text{HeII}=0.0$). The transfer function value in the region encircled by the black dashed line is negative and the $\log_{10}$ of the absolute value is shown. The third panel shows the error in $\log_{10}\Pbg$ (defined as $\Delta \log_{10} \Pbg \equiv\log_{10}|\Pbg_\text{DNN}|-\log_{10}|\Pbg_\text{table}|$), which is concentrated on the physical features of the transfer functions. The fourth panel shows the relative error in the energy transfer function (see more details in Sec.~\ref{subsec:tfvaluepred}). The errors are concentrated on the diagonal that corresponds to photon propagation and redshifting.}
\label{fig:comptf_hep}
\end{figure*}

\paragraph{Redshift regimes and matter ionization dependence.}
All photon transfer functions depend on the redshift, as the photon cooling processes involve interactions with the redshift-dependent photon background and/or interstellar medium.
For late redshifts ($z<40$) encompassing the epoch of reionization, the transfer functions are allowed to vary with both the ionized hydrogen fraction $x_\text{HII}$ and the singly ionized helium fraction $x_\text{HeII}$, which can be altered by exotic energy injections.
For redshifts between helium recombination and reionization ($40<z<1600$), the ionized helium fraction can be safely approximated as zero \cite{liu2020darkhistory}; the transfer functions are pre-computed assuming no helium ionization, but can depend on the hydrogen ionized fraction.
Before helium recombination ($z>1600$), exotic energy injections consistent with current experimental bounds have little impact on the thermal equilibrium determining hydrogen and helium ionization levels \cite{liu2020darkhistory}, so \dhis uses \texttt{RECFAST} \cite{wong2008well} ionization fractions as a baseline to pre-compute the transfer functions, which only depends on redshift. 

For the DNN implementation, the flexibility of the network allows one DNN to be trained on the entire redshift range $4<z<3000$ with a learned dependence on the ionization levels, for each transfer function. However, using different networks for different redshift regimes gives slightly better accuracy, and the latter approach is chosen in this version of \dhis.

\paragraph{Redshift step coarsening and energy conservation.}
In the previous version of \dhis, photon transfer functions are computed with redshift step $\Delta\log(1+z)=0.001$. One can choose to increase the (log) step size to multiples of 0.001 to speed up computation. To combine multiple redshift steps, \dhis pre-composes multiple propagation transfer functions $\overline{\mathbf{P}}_\gamma$ and applies them appropriately to the deposition transfer functions $\overline{\mathbf{D}}_\gamma$, $\overline{\mathbf{D}}_e$, and $\overline{\mathbf{D}}_c^\text{high}$. Since the DNN implementation of transfer functions introduces a small amount of error that can accumulate over many redshift steps, in order to increase numerical stability,
we train the DNNs to reproduce the pre-composed transfer functions, and require the use of a fixed log redshift step of $\Delta\log(1+z)=0.012$.
\footnote{It was shown in \cite{liu2020darkhistory} that an increased redshift step size at this level will lead to errors smaller than $10^{-3}$, and so will not be the primary contributor to numerical error in our context. Note also that using a $n$-fold coarsened log redshift step is not equivalent to using every $n^\text{th}$ entry in the transfer function table. The redshift abscissa used for the table are not directly related to (and in general are coarser than)
the redshift steps used during a run; the table is interpolated to obtain the transfer functions at each redshift step.} To further decrease numerical error, we store the total fraction of the injected energy entering each type of secondary particle spectrum, for an injected photon of any given energy, and use these data (which are of $\sim1\%$ the size of the transfer functions) to ensure energy conservation while accounting for energy loss to redshifting. We discuss the precise procedure of transfer function pre-composing and imposing energy conservation in Appendix~\ref{appd:coarsening}.

\begin{figure*}[ht]
\includegraphics[clip, width=\textwidth]{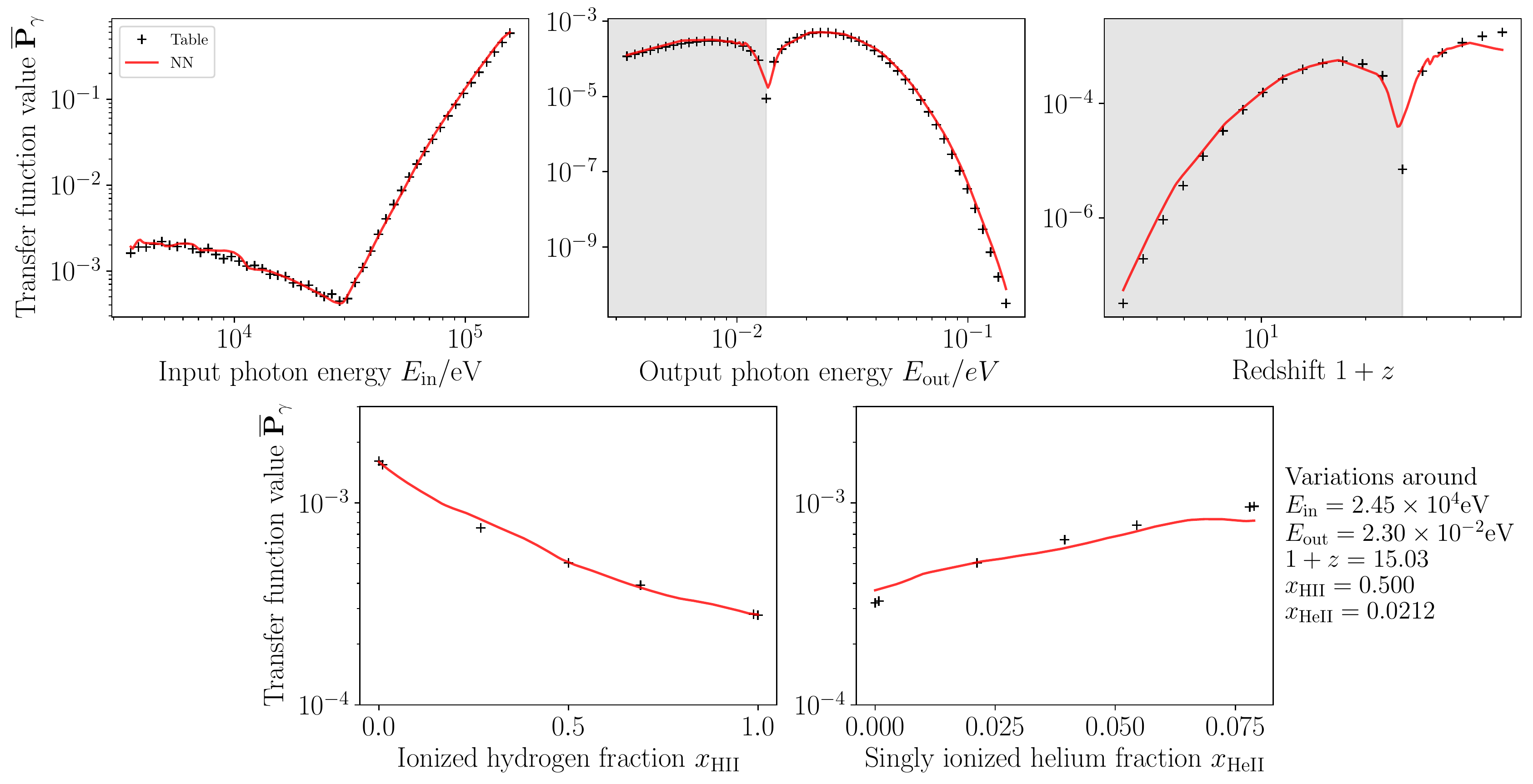}
\caption{\textbf{The DNN high energy photon transfer function network's fitting and interpolation of the table transfer function values.} The panels show the DNN transfer function's (redshift regime $4<z<40$) dependence on each of the five input variables, in a limited region around a central value indicated on the plot. The shaded region is where both the table value and the DNN prediction are negative (and the log of the absolute value is shown). A fairly consistent relative error below $10\%$ is observed.}
\label{fig:interp}
\end{figure*}

\section{Transfer Functions from Neural Networks}
\label{sec:nntf}

As described earlier and as indicated in Fig.~\ref{fig:model}, we update the largest transfer function tables (\texttt{ics\_thomson}, \texttt{ics\_rel}, \texttt{ics\_engloss}, \texttt{highengphot}, and \texttt{lowengelec}) to DNN networks that take in input and output particle energies, redshift, and possibly (depending on redshift) the ionized hydrogen fraction and singly ionized helium fraction, to produce the transfer function value $P$. In general, $P$ can vary greatly across many orders of magnitude. (For example, after recombination the probability for a 10 keV photon to free stream, losing energy only through redshifting, is substantial, while the chance of it directly producing secondary photons of 1 keV is very close to 0, since this outcome is not kinematically allowed in a single Compton scattering event, nor is the scattered electron produced by Compton scattering or photoionization able to up-scatter other photons to 1 keV.) As such, we train the networks to output the natural logarithm of the transfer function value $\log P$.
Similarly, since the input/output energy abscissa and our redshift steps are also binned in log space by default, the networks also take the log value of these as inputs. The ionized hydrogen fraction and singly ionized helium fraction are linearly scaled to match the spread of other parameters before being fed into the DNN.

For high energy photons, the redshift-coarsened transfer functions (see Sec.~\ref{subsec:photontf}) are fitted. Note also that the transfer function value $P$ can be negative since by convention the CMB spectrum is subtracted from the transfer function and the negative values (and the positive values at higher energies) represent CMB photons being upscattered. In this case $\log|P|$ is predicted by the DNNs, and the negative value region is recovered in a post-processing step that identifies local minima of $\log|P|$ (which takes up an negligible time compared to the rest of the \dhis routine). Additionally, the transfer functions values near the diagonal in the input/output energy dimensions (corresponding to the free-streaming photons) are adjusted to enforce energy conservation up to redshifting. For details please refer to Appendix \ref{appd:coarsening}.

After adjusting hyperparameters,
we find that for all transfer functions in question, it is sufficient to use DNNs with 7 hidden layers with 400 neurons each, making the number of parameters per DNN $\sim 400^2 \times (7 - 1)\sim9.7\times10^5$,
and $2.9\times10^6$ for each transfer function built from 3 such networks.

Training is done with TensorFlow 2.0~\cite{tensorflow2015-whitepaper} and Keras~\cite{chollet2015keras}; Adagrad~\cite{lydia2019adagrad} is used as the optimizer, with the mean squared error of $\log P$ as the loss function. Each DNN is trained on 2 V100 GPUs for $\order{10}$ hours or equivalent. For each epoch, training data is generated by interpolating the multi-dimensional transfer function table on uniformly random sampled inputs. Since training data is not reused across epoch, there is no concern of overfitting to a fixed subset of full data set. Trainings are terminated with the evaluation after each iteration stops improving significantly. To check for systematic offsets of the table transfer functions and the NNs, we trained multiple NNs with random initial values and randomly sampled training data. We found no obvious systematic offsets between the ensemble of NNs and the tables; e.g. at any given point the different NNs both underpredicted and overpredicted the table data.

The codes related to the DNN transfer functions are stored in the \texttt{nntf} module under \dhis. A new example file ``Example 12: Using Neural Network transfer functions.ipynb"\href{https://github.com/hongwanliu/DarkHistory/blob/nntf_alt_hep_clean/examples/Example_12_Using_Neural_Network_Transfer_Functions.ipynb}{\faGithub} is provided to demonstrate using the DNN transfer functions and comparison with the baseline \dhis (which will be available if the appropriate data tables are present).

\section{Performance}
\label{sec:performance}

In this section, we describe the accuracy and speed of generation of the DNN transfer functions, as well as the accuracy of full runs over a range of simple DM injection scenarios using the DNN transfer functions.

\subsection{Transfer function value prediction}
\label{subsec:tfvaluepred}

In Fig.~\ref{fig:comptf_hep}, a high energy photon transfer function generated by a DNN is compared against one interpolated from tables. As one can see, the errors in the raw output i.e. logarithm value of the transfer functions are concentrated near the distinct physical features in the transfer function, such as at the output photon energy of $\sim0.5$~MeV corresponding to positronium decay. The absolute values in logarithm errors $\Delta\log_{10} |P|$ can be interpreted as relative errors in $|P|$ (up to a $\ln 10$ factor). In this particular slice through the high energy photon transfer function, the average $\Delta\log_{10}|P|$ when $|P|>10^{-20}$ is 0.017, corresponding to a relative error of $\sim4$\%. (The range of $\log_{10}|P|$ is about $\sim-45$ to 6). Overall $\Delta\log_{10} |P|$ is comparable to this value, for all DNN transfer functions. A summary of the errors can be found in Tab.~\ref{tab:tf_err_time}.

\begin{figure*}[ht]
\includegraphics[clip, width=\textwidth]{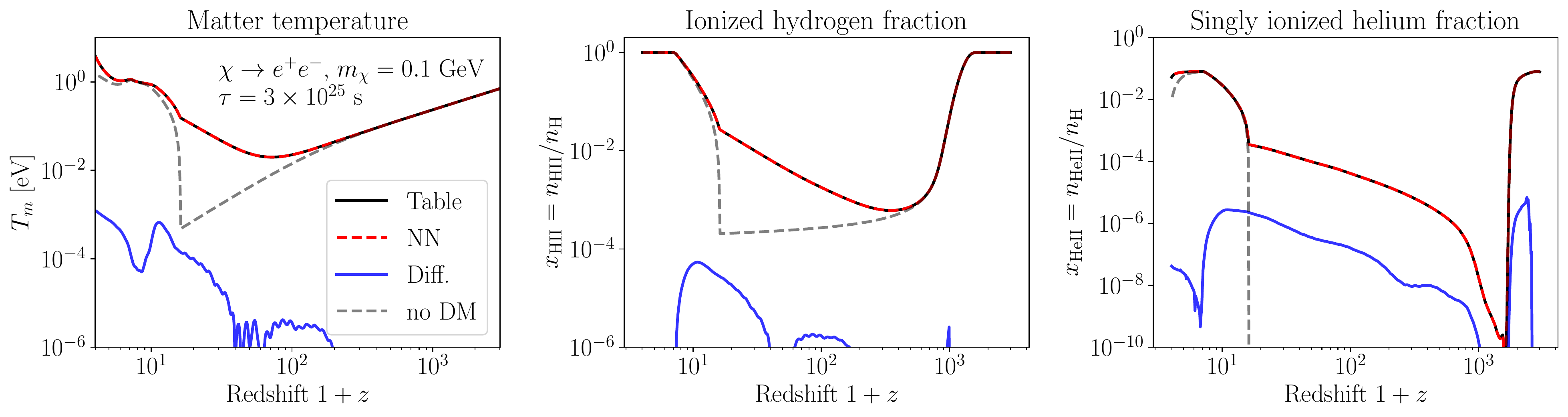}
\caption{\textbf{Example evolution of matter temperature and ionization levels.} The evolution of matter temperature and ionization levels, in an example scenario where 0.1~GeV DM is decaying into $e^+e^-$ pairs, as predicted using the original tabulation ({\it black lines}) and using the DNN transfer functions ({\it red dashed lines}). The blue line shows the difference between the two, and the gray dashed line shows the standard evolution without any external injections.}
\label{fig:TmxHxHe}
\end{figure*}

To see the impact of these errors in a \dhis evolution run, it is also useful to look at errors in energy (per bin) transition rates, besides the particle number transition rates. The energy transfer function de-emphasize errors at low energies where many particles can be produced with only a small fraction of the original particle's energy, and thus such errors have a small effect on heating and ionization. Since the standard photon transfer functions are maps between particle number spectra $N_i$, the transfer function values have the physical meaning of number transition rates. The particle energy spectrum is related to the number spectrum by
\begin{equation}
    E_i = \mathcal E_i N_i=\mathcal E_i^2\left(\frac{\dd N}{\dd\mathcal E}\right)_i,
\end{equation}
where $N_i$ is the number of particles in the $i$-th bin, and $\mathcal E_i$ its central energy. (Note that the energy bins are log-spaced.) For a particle number transfer function $P$ with
\begin{equation}
    N^\text{out}_i = \sum_j N^\text{in}_j P_{ji},
\end{equation}
the corresponding energy transfer function $P^E$ is defined by
\begin{equation}
    E^\text{out}_i = \sum_j E^\text{in}_j P_{ji}^E
    \implies P_{ji}^E = P_{ji}\mathcal E_i/\mathcal E_j.
\end{equation}
In the last panel of Fig.~\ref{fig:comptf_hep}, we show the relative error in the high energy photon transfer function. As expected, the errors are concentrated on the highest output energy for a given input energy. The relative errors are generally sub-percent.

Fig.~\ref{fig:interp} shows that the DNN networks interpolate sensibly between the fixed abscissa values in the transfer function tables, for the high energy photon transfer function in the lowest redshift regime ($4<z<40$), as an example. The errors in the transfer function interpolation are consistent with average values shown in Tab.~\ref{tab:tf_err_time}.

The time it takes to generate a photon transfer function is about or less than 1 second on an 8-CPU personal computer. ICS transfer functions that are only generated once per run takes only slightly longer at $\lesssim3$~s. The accuracy and prediction time for all other transfer functions can be seen in Tab.~\ref{tab:tf_err_time}.

\begin{figure*}[ht]
\includegraphics[clip, width=0.8\textwidth]{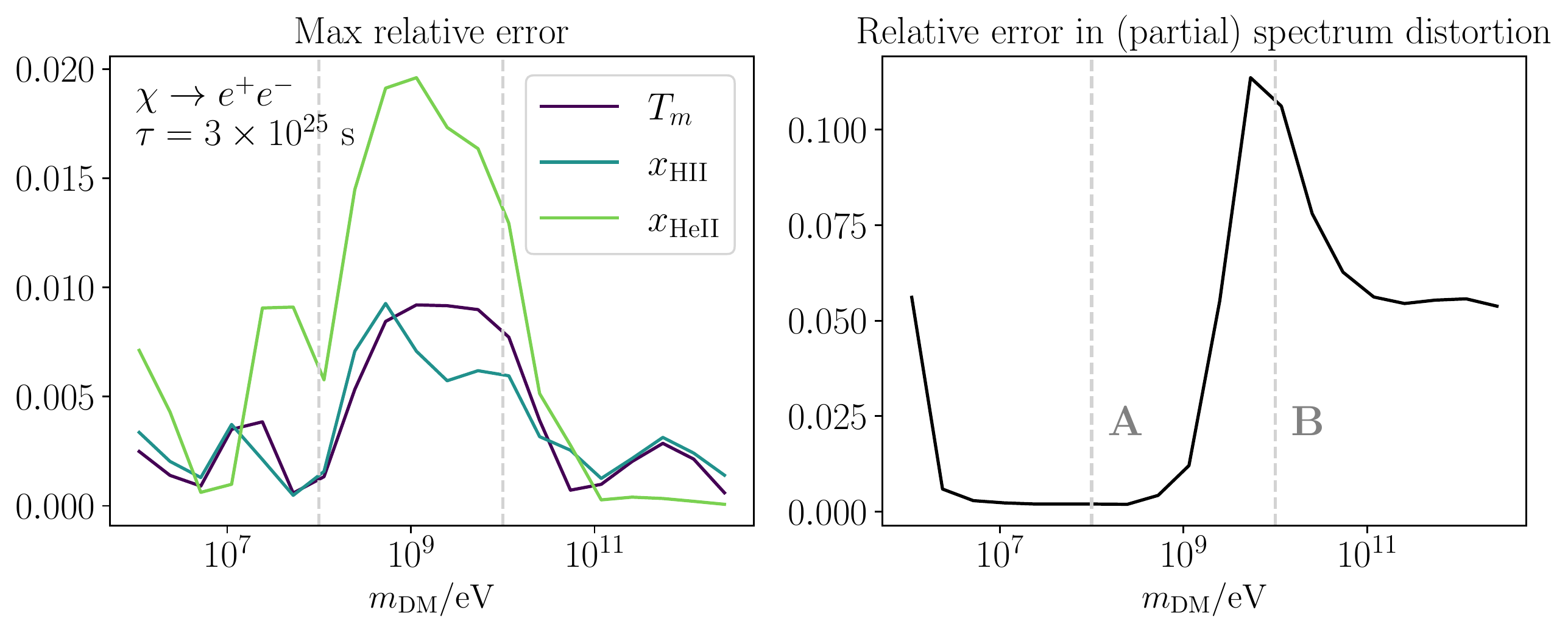}
\caption{\textbf{Relative errors in temperature, ionization levels, and low energy photon distortion across a range of DM mass in a scenario with DM decaying to $e^+e^-$.} The left panel shows the maximum relative error at any point in the evolution in $4<z<3000$ of matter temperature, ionized hydrogen fraction, and singly ionized helium fraction across a range of DM exotic electron injection energies*. (*Due to its very small absolute value, the relative error for singly ionized helium fraction $x_\text{HeII}$ when $n_\text{HeII}/n_\text{He}<10^{-3}$ is not included in this plot. Since $n_\text{HeII}/n_\text{He}$ changes rapidly between $\order{1}$ and $<10^{-3}$, we are essentially only counting its relative error when its value is order unity.) The relative errors are generally below 5\%. The right panel shows the maximum error over the maximum value in the low energy photon spectral distortion at $z=0$. The spectral distortion errors at the two dashed-line values of $m_\text{DM}=0.1$~GeV and $10$~GeV are shown in Fig.~\ref{fig:spec}. }
\label{fig:scan_errors_decay_elec}
\end{figure*}

\begin{figure*}[ht]
\includegraphics[clip, width=0.8\textwidth]{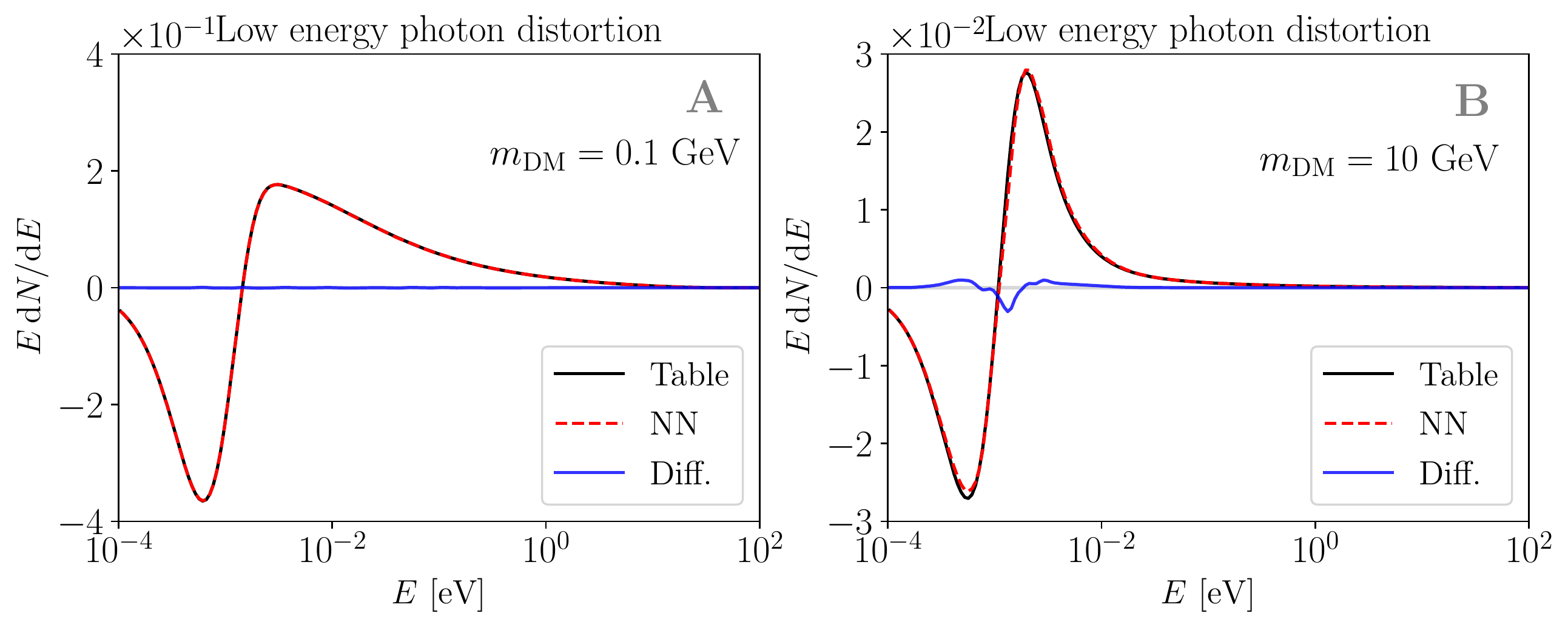}
\caption{\textbf{Example partial low-energy photon spectral distortion in the present day.} The two panels show two examples of low energy photon spectral distortion outputs (see definition in Sec.~\ref{sec:tf}) from two different runs: 0.1~GeV and 10~GeV DM decaying to $e^-e^+$ pairs. They represent the extremes of large and small relative errors for runs over the full range of DM masses we consider, as shown in Fig.~\ref{fig:scan_errors_decay_elec}. The photon number density (per bin) is normalized against the baryon number density.
The black lines represent outputs generated using tabulated transfer functions while the red dashed lines represent that using the DNN transfer functions. The blue line shows the different between the two. Note that the relatively large errors shown in the right panel are in part due to the error in the location of the zero.}
\label{fig:spec}
\end{figure*}

\subsection{Performance over a range of scenarios}

Fig.~\ref{fig:TmxHxHe} shows the evolution of integrated variables in one particular setting: 0.1~GeV DM particles decaying into electron-positron pairs (this is the same as the example used in Fig. 4 of Ref.~\cite{liu2020darkhistory}). In this example, the error introduced by the DNN in the matter temperature history and hydrogen ionization history is consistently sub-percent, while the error in singly ionized helium fraction is sub-percent when $n_\text{HeII}/n_\text{He}>10^{-3}$. For a scenario where DM decays into electron-positron pairs, over a range of DM rest masses, the maximum relative error for temperature and ionized hydrogen fraction over the entire redshift range is consistently below 2\%, as shown in Fig.~\ref{fig:scan_errors_decay_elec}. Taking into account injection scenarios with DM decaying to photons, and also undergoing $s$-wave annihilation into $e^+e^-$ or photons, the relative error is always below 8\%. \dhis also computes the partial photon spectral distortion from high energy photon and electron processes, mostly from ICS of electrons or positrons on the CMB. This distortion is stored in the low energy photon spectrum. Note that spectral distortions arising from atomic transitions, due to photons and electrons below 3~keV, are not included in this spectrum (which is why we label it as ``partial'' or ``incomplete''). The DNN transfer functions introduce a small amount of error in this spectral distortion, as shown in Fig.~\ref{fig:spec} and Fig.~\ref{fig:scan_errors_decay_elec}. While the shape of the spectral distortion is generally correct, the error in the location of the distortion zero can cause errors with a magnitude up to 10\% of the distortion magnitude due to the errors in the distortion zero location. In a future update of \dhis, we are anticipating an update to include the correct treatment of the complete photon spectral distortion including contributions from atomic transitions. A small photon spectral error would allow the DNN functions to be used simultaneously with these updates.

In Appendix \ref{appd:errors}, we include the errors in some other exotic injection scenarios, including DM decaying to photon pairs, and DM annihilating to photon or $e^+e^-$ pairs. Although these examples cover only a few simple injection scenarios, they can serve to test the whole range of \dhis's dependence on transfer functions, since all exotic energy injections are converted to either $e^+e^-$ or photon injections. We expect exotic energy injections with a more complicated injection spectrum (such as e.g. annihilation to quarks) to have similar amounts of error associated with using the DNN transfer function.

\begin{figure*}[ht]
\includegraphics[clip, width=0.8\textwidth]{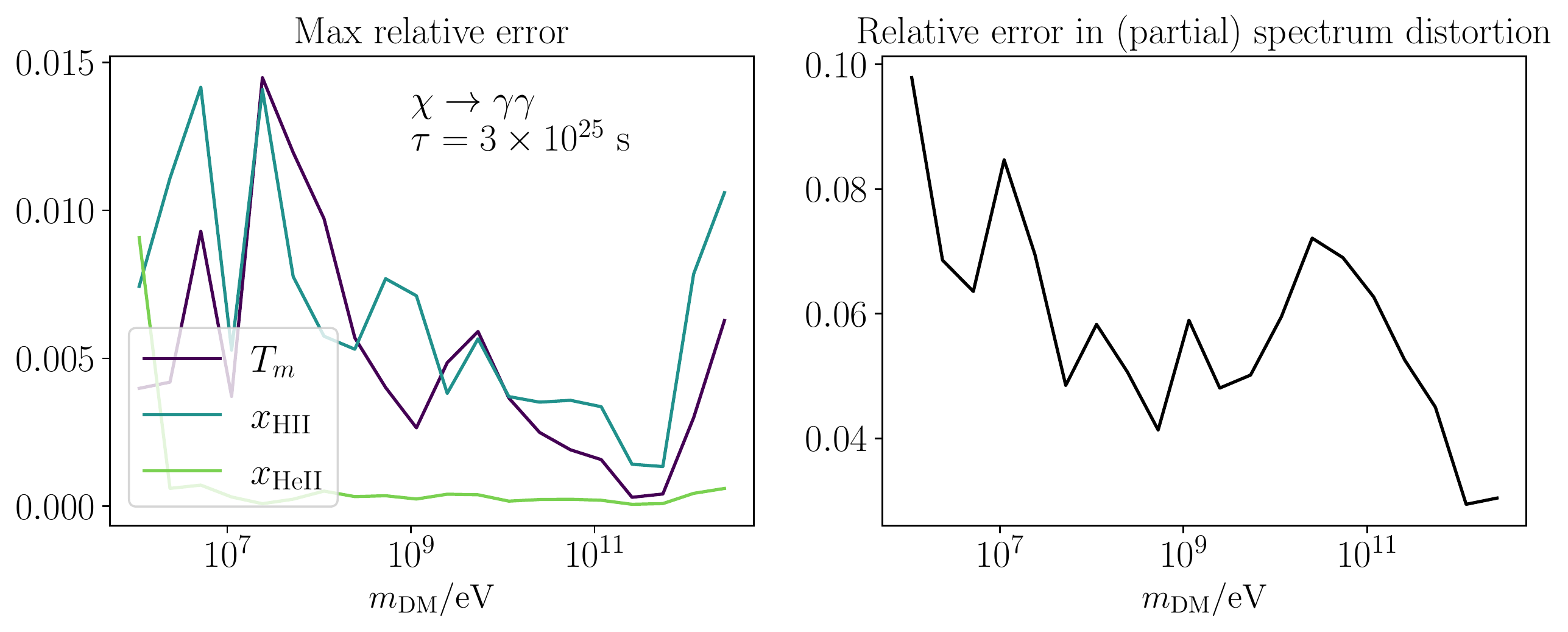}
\caption{\textbf{Relative errors in physical quantities across a range of DM mass, in a scenario with DM decaying to a pair of photons.}
The panels have similar construction to Fig.~\ref{fig:scan_errors_decay_elec}. Note that the matter temperature and ionization fractions have max relative errors below 2\%, and the errors in the low-energy photon spectral distortion are also below 10\%.}
\label{fig:scan_errors_decay_phot}
\end{figure*}

\section{Conclusion}
\label{sec:conclusion}

In this work, we have made use of simple Dense Neural Networks to approximate complex and multi-dimensional transfer functions in \dhis to reduce storage and memory usage, as well as to enable the possibility of adding more parameter dependence to the transfer function. The DNN transfer functions achieve good accuracy in computing the evolution history of matter temperature and ionization, as well as the partial CMB spectral distortion evaluated by the current version of \dhis; typical errors are at the few percent level, and are comparable to or smaller than estimates of systematic uncertainties in previous studies of constraints on energy injection \cite{galli2013systematic,Slatyer:2015jla,liu2020darkhistory}.

The DNN-based functionality is available in the \dhis Github repository at \url{https://github.com/hongwanliu/DarkHistory}, and the necessary data files (one can choose to download the large tables, or DNN and auxiliary files, or both) are hosted on \href{https://doi.org/10.5281/zenodo.6819281}{Zenodo} (see the Github repository for details).

While the use of DNNs offers one solution to this challenge, there may well be other viable solutions. The information in the DNNs still seems likely to be an over-representation of the piecewise-smooth transfer functions. We have briefly explored some alternative methods, including fitting to conventional functions directly, and with the assistance of symbolic regression techniques \cite{schmidt2009distilling}; however, DNN networks stand out as the best solution (so far) in terms of fitting accuracy and ease of implementation. There is work ongoing to expand the capabilities of \dhis, and we look forward to exploring NN-based and alternative techniques in this context.

\section*{Appendix}

\subsection{Redshift step coarsening and energy conservation}
\label{appd:coarsening}

In this Appendix, we describe how the photon transfer function change with coarsened redshift step, (expanding on Sec.~III.E.3 of Ref.~\cite{liu2020darkhistory}), and how energy conservation while correctly accounting for photon energy loss due to redshift is implemented.

\subsubsection{Transfer functions without coarsening}
In \dhis's \texttt{main.evolve} function, evolution is discretized into log-normal redshift steps with fixed spacing $d$ (\texttt{dlnz} in code) where the next redshift $z'$ is expressed in terms of $z$ such that:
\begin{equation}
    \log(1+z')=\log(1+z)-d.
\end{equation}

Following \dhis's flow described in Fig.~\ref{fig:dhtfs}, to obtain the propagating photon spectrum $\Npropg$, low energy photon spectrum $\Nlowg$, low energy electron spectrum $\Nlowe$ and energy deposition array $\Ec$, we retrieve the corresponding transfer functions $\Pbg$, $\Dbg$, $\Dbe$, and $\Dbc$ at a middle redshift $z_\text{mid}$ given by
\begin{equation}
    \log(1+z_\text{mid})=\log(1+z)-d/2.
\end{equation}
The corresponding transfer functions $\Pbg$, $\Dbg$, and $\Dbe$ all take in the propagating photon spectrum $\Npropg$ at redshift $z$ and produce secondary spectra at $z'$. ($\Dbc$ produces energy deposition in this redshift step.)
They are implemented as:
\begin{equation}
\label{eq:applytf}
\begin{aligned}
    \Npropg(z')&=\Npropg(z)\cdot\Pbg(z_\text{mid})=\sum_j\Npropg_j\Pbg_{ji} \\
    \Nlowg(z')&=\Npropg(z)\cdot\Dbg(z_\text{mid}) \\
    \Nlowe(z')&=\Npropg(z)\cdot\Dbe(z_\text{mid}) \\
    \Ec(z')&=\Npropg(z)\cdot\Dbc(z_\text{mid})=\sum_j\Npropg_j\Dbc_j,
\end{aligned}
\end{equation}
where $i$ and $j$ are indices of discretized energy abscissa.

\begin{figure*}[ht]
\includegraphics[clip, width=0.8\textwidth]{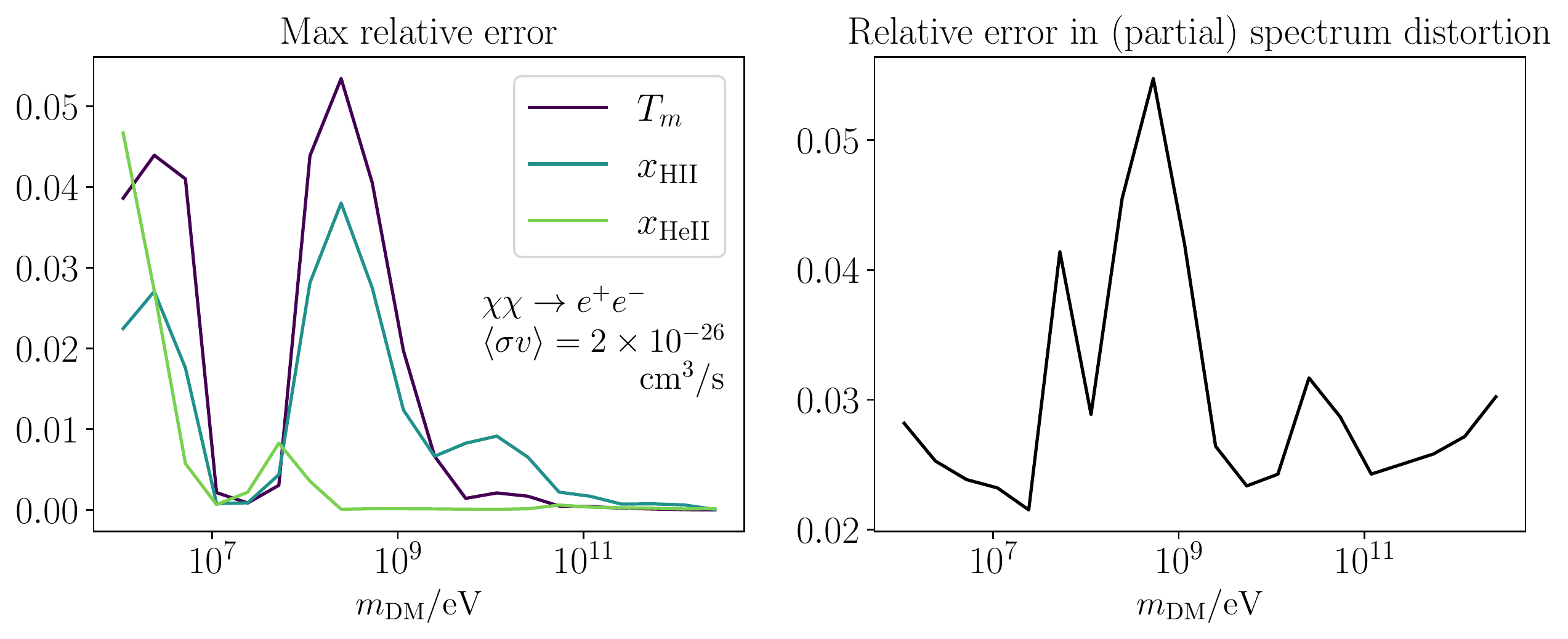}
\caption{\textbf{Relative errors in physical quantities across a range of DM mass in a scenario where DM undergoes $s$-wave annihilation to a electron/positron pair.}
The panels have similar construction to Fig.~\ref{fig:scan_errors_decay_elec}.
}
\label{fig:scan_errors_swave_elec}
\end{figure*}

\begin{figure*}[ht]
\includegraphics[clip, width=0.8\textwidth]{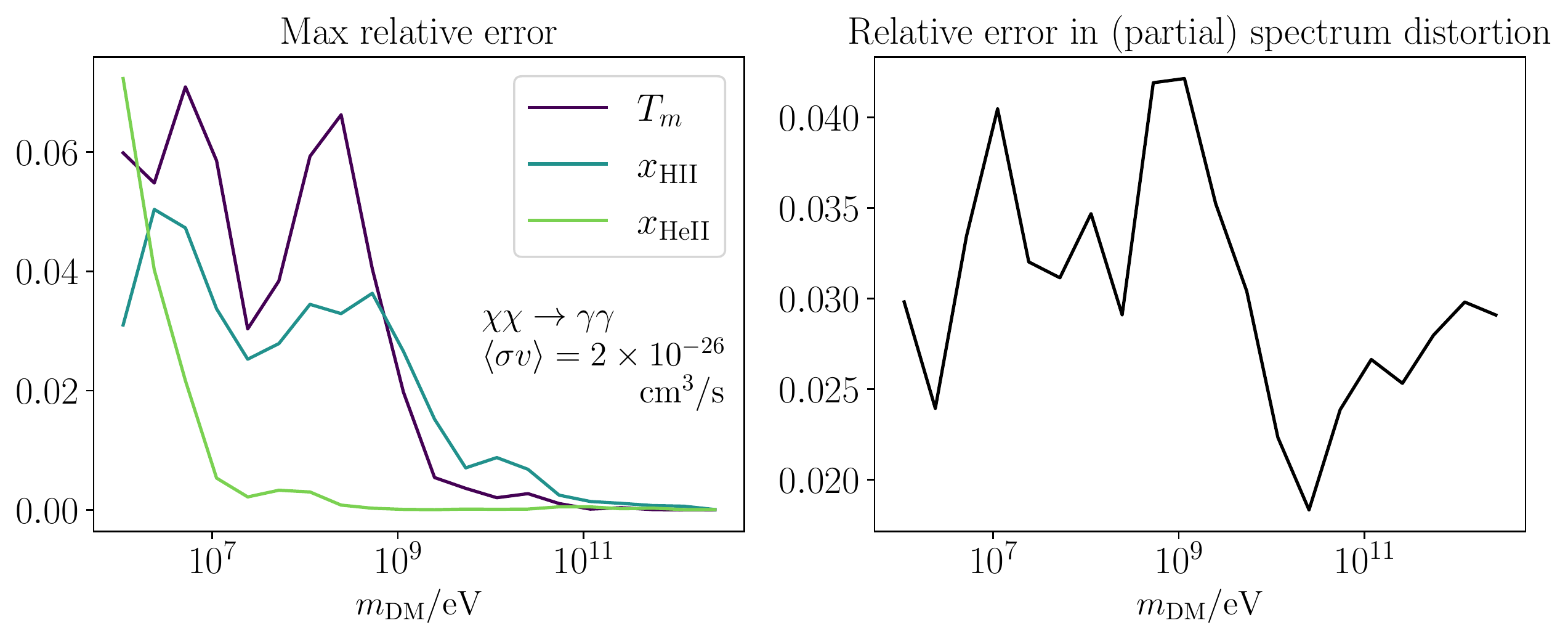}
\caption{\textbf{Relative errors in physical quantities across a range of DM mass in a scenario where DM undergoes $s$-wave annihilation to a photon pair.}
The panels have similar construction to Fig.~\ref{fig:scan_errors_decay_elec}.
}
\label{fig:scan_errors_swave_phot}
\end{figure*}

Energy conservation can be enforced straightforwardly: for an injection in any photon energy bin with central energy $\Ei$, the total output energy on the right hand side of the above equations should add up to $\Ei$ minus the loss to redshifting of the propagating photons. Photons below a certain energy $E_\text{relevant}$ do not contribute to redshift energy loss because they either dump all of their energies efficiently within one redshift step or free stream and no longer interact, in which case they are stored as an array history of low energy photons $\Nlowg(z')$) and not immediately redshifted.

For the propagating photons that are redshifted by $\Pbg$, the energy lost is approximately:
\begin{equation}
\begin{aligned}
    |E_{\text{redshift},i}|=\left(1-\frac{1+z'}{1+z}\right)\Ei&=\frac{z-z'}{1+z}\,\Ei\approx d\,\Ei\\
    &\text{for }\Ei>E_\text{relevant}.
\end{aligned}
\end{equation}
Let the energy abscissa (log-central energies of each bin) for photons and electron be $\Eg_i$ and $\Ee_i$ respectively. We can express the above equation as
\begin{equation}
    |E_{\text{redshift},i}|=d_i\Ei
\end{equation}
where $d_i=d$ when the photon with energy $\Ei$ should be redshifted and $d_i=0$ otherwise. (This renders $d_i$ a function of redshift and hydrogen and helium ionization levels in general.)

Then the energy conservation constraint can be written as
\begin{equation}
    \Ei=\sum_j\left(\Pbg_{ij}\Eg_j+\Dbg_{ij}\Eg_j+\Dbe_{ij}\Ee_j+\Dbc_i+d_i\Ei\right)
\end{equation}
This relation is imposed by shifting the near diagonal (propagating) part of the high energy photon transfer function $\Pbg$. Any energy non-conservation due to numerical errors or errors from approximating the transfer function as DNNs can be absorbed into this shift.

\begin{table*}[ht]
\centering
\begin{tabular} {|c||c|c||c|c|}
\hline\hline
DNN transfer function & $|\Delta\log_{10}|P||$ & Prediction time & Table size & NN size \\
\hline
high energy photon (compounded) regime 0 & 0.029 & 1.01 s & & \\
high energy photon (compounded) regime 1 & 0.043 & 1.04 s & 4.7Gb & 11.4Mb \\
high energy photon (compounded) regime 2 & 0.016 & 1.02 s & & \\ \cline{4-5}
high energy photon (propagator) regime 0 & 0.029 & 1.02 s & & \\
high energy photon (propagator) regime 1 & 0.043 & 1.05 s & 4.7Gb & 11.4Mb \\
high energy photon (propagator) regime 2 & 0.062 & 1.07 s & & \\ \cline{4-5}
low energy electron regime 0 & 0.047 & 0.382 s & & \\
low energy electron regime 1 & 0.049 & 0.376 s & 4.7Gb & 11.4Mb \\
low energy electron regime 2 & 0.040 & 0.370 s & & \\ \cline{4-5}
ICS Thomson              & 0.00199 & 2.82 s & 0.93Gb & 3.8Mb \\
ICS relativistic         & 0.00410 & 2.12 s & 0.93Gb & 3.8Mb \\
ICS electron energy loss & 0.00250 & 2.30 s & 0.93Gb & 3.8Mb \\
\hline
\end{tabular}
\caption{\textbf{Error, prediction time, and size of DNN transfer functions used in \texttt{DarkHistory}.} The error of the DNN transfer functions in comparison to the tabular transfer functions is defined as $|\Delta\log_{10}|P||\equiv|\log_{10}|P_\text{DNN}|-\log_{10}|P_\text{table}||$ when $|P|>10^{-20}$. Both the errors and prediction times are evaluated from random draws in the relevant domains for each DNN until the values stabilize. The prediction times are evaluated on a 8-CPU personal computer.
All DNN transfer functions use 6 hidden layers of 400 neurons each, but the evaluation time differs due to the length of the input array corresponding to the size of the evaluated transfer function matrix ($\sim10^5$ for high energy photon, $\sim4\times10^4$ for low energy electron, $\sim2.5\times10^5$ for ICS).}
\label{tab:tf_err_time}
\end{table*}

\subsubsection{Coarsening}
\dhis can enlarge the redshift step $d$ to an multiple of a preset step $d_0$. Let the coarsening multiple be $c$, then the next redshift step $z'$ to $z$ is such that
\begin{equation}
    \log(1+z')=\log(1+z)-c\cdot d_0.
\end{equation}
The photon and electron transfer functions are built with log redshift step $d_0$, so we have to reconstruct the transfer functions for $d$. We first obtain single-step transfer functions (and $d_i$) with current ionization levels and a redshift value $z_\text{mid}$ in the middle of the large redshift step $d$:
\begin{equation}
    \log(1+z_\text{mid})=\log(1+z)-c\cdot d_0/2.
\end{equation}
Then, we approximate the large redshift step as consisting of $c$ single redshift steps, each with the same transfer function applied. The compounded transfer functions can be expressed as
\begin{equation}
\begin{aligned}
    \Npropg(z')&=\Npropg(z)~\Pbg^c \\
    \Nlowg(z')&=\Npropg(z)\left(1+\Pbg+\Pbg^2+\cdots+\Pbg^{c-1}\right)\Dbg \\
    \Nlowe(z')&=\Npropg(z)\left(1+\Pbg+\Pbg^2+\cdots+\Pbg^{c-1}\right)\Dbe \\
    \Ec(z')&=\Npropg(z)\left(1+\Pbg+\Pbg^2+\cdots+\Pbg^{c-1}\right)\Dbc, \\
\end{aligned}
\end{equation}
with all transfer function evaluated at $z_\text{mid}$. In the DNN implementation, the compounded transfer functions $\Pbg^c$, $\left(1+\Pbg+\Pbg^2+\cdots+\Pbg^{c-1}\right)$, and $\Dbe$ are learned as DNN networks, and this step can be carried out without using the value of $\Pbg$ itself. (The low energy photon transfer function $\Dbg$ can be quickly reconstructed using the CMB energy loss information.)

Imposing energy conservation is similar: At each $d_0$ step, the redshift energy loss is $\Npropg(z+nd_0)\cdot d_{0i}\Ei$, where $\Npropg(z+nd_0)=\Npropg(z)\cdot\Pbg^n$. So the total redshift energy loss for a $\Ei$ photon is
\begin{equation}
     |E_{\text{redshift},i}|=\left[\left(1+\Pbg+\Pbg^2+\cdots+\Pbg^{c-1}\right) d_0\mathcal E\right]_i.
\end{equation}
Let $\Sbg=1+\Pbg+\Pbg^2+\cdots+\Pbg^{c-1}$. Then the energy conservation condition is
\begin{equation}
\begin{aligned}
    \Ei=\Pbg^c_{ij}\Eg_j+\Sbg \Big( \Dbg_{ij}\Eg_j+\Dbe_{ij}\Ee_j&+\Dbc_i+d_i\Ei \Big) \\
    &\forall i\text{, sum on }j.
\end{aligned}
\end{equation}
Again, the propagating photon spectrum can be adjusted to account for energy non-conservation from numerical and DNN prediction errors.

\subsection{Performances in other DM injection scenarios}
\label{appd:errors}

Performance metrics of \dhis using DNN transfer functions are further demonstrated in Fig.~\ref{fig:scan_errors_decay_phot} (DM decaying to photon pairs), Fig.~\ref{fig:scan_errors_swave_elec}, (DM annihilating to electron/positron pairs with $s$-wave cross section), and Fig.~\ref{fig:scan_errors_swave_phot} (DM annihilating to photon pairs with $s$-wave cross section).

\subsection{Performances and prediction time of individual DNN transfer functions.}
\label{appd:tferrors}
The prediction accuracy and prediction time of individual DNNs are summarized in Tab.~\ref{tab:tf_err_time}.

\section*{Acknowledgements}
It is a pleasure to thank Hongwan Liu, Wenzer Qin, Gregory Ridgway, Jesse Thaler, Ken van Tilburg, Kathrin Nippel, Nils Schoeneberg, Anastasia Fialkov, Stefan Heimersheim, Julian Mu\~noz, Laura Lopez Honorez, Vivian Poulin, and Yacine Ali-Haimoud for useful conversations pertaining to this work. YS and TRS were supported by the U.S. Department of Energy, Office of Science, Office of High Energy Physics of U.S. Department of Energy under grant Contract Number DE-SC0012567 through the Center for Theoretical Physics at MIT, and the National Science Foundation under Cooperative Agreement PHY-2019786 (The NSF AI Institute for Artificial Intelligence and Fundamental Interactions, \url{http://iaifi.org/}).
The computations in this paper were run on MIT LNS's Erebus machine and the FASRC Cannon cluster supported by the FAS Division of Science Research Computing Group at Harvard University.
This research made use of the
\texttt{IPython}~\cite{PER-GRA:2007}, \texttt{Jupyter}~\cite{Kluyver2016JupyterN}, \texttt{matplotlib}~\cite{Hunter:2007},  \texttt{NumPy}~\cite{harris2020array}, \texttt{Tensorflow}~\cite{tensorflow2015-whitepaper}, \texttt{Keras}~\cite{chollet2015keras}, and \texttt{SciPy}~\cite{2020SciPy-NMeth} software packages.
\bibliography{main}

\end{document}